\documentclass[aps,prl,reprint,showpacs,showkeys,superscriptaddress,preprintnumbers,groupedaddress]{revtex4-1}
\usepackage{amsmath,amssymb,bm,mathrsfs}
\usepackage{srcltx}
\usepackage{dsfont}
\usepackage{epsfig}
\usepackage{slashed}
\usepackage{bbold}
\usepackage{psfrag}
\usepackage{xcolor}
\PassOptionsToPackage{caption=false}{subfig}
\usepackage{subfig}
\usepackage{xfrac}
\usepackage{multirow}
\usepackage{booktabs}
\usepackage[colorlinks=true,linkcolor=blue,citecolor=blue,urlcolor=violet]{hyperref}

\newcommand{\be}{\begin{equation}}
\newcommand{\ee}{\end{equation}}
\newcommand{\bea}{\begin{eqnarray}}
\newcommand{\eea}{\end{eqnarray}}

\newcommand{\nn}{\nonumber}
\def\({\left(}
\def\){\right)}

\usepackage{soul}

\begin{document}

\title{A Fourth Exception in the Calculation of Relic Abundances}

\author{Raffaele Tito D'Agnolo}
\email{raffaele.dagnolo@epfl.ch}
\affiliation{Theoretical Particle Physics Laboratory, Institute of Physics, EPFL, Lausanne, Switzerland} 
\affiliation{Institute for Advanced Study, Princeton, NJ 08540, USA.}   

\author{Duccio Pappadopulo}
\email{duccio.pappadopulo@gmail.com}
\affiliation{
Center for Cosmology and Particle Physics,
Department of Physics, New York University, New York, NY 10003, USA.
} 

\author{Joshua T. Ruderman}
\email{ruderman@nyu.edu}
\affiliation{
Center for Cosmology and Particle Physics,
Department of Physics, New York University, New York, NY 10003, USA.
}

\begin{abstract}
We propose that the dark matter abundance is set by the decoupling of inelastic scattering instead of annihilations.  This {\it coscattering} mechanism is generically realized if dark matter scatters against
states of comparable mass from the thermal bath.  Coscattering points to dark matter that is
exponentially lighter than the weak scale and has a suppressed annihilation rate, avoiding stringent constraints from indirect detection.  
Dark matter upscatters into states whose late decays can lead to observable distortions to the blackbody spectrum of the cosmic microwave background.

 \end{abstract}

\maketitle

\noindent {\bf  Introduction:}\
Dark Matter (DM) constitutes most of the matter in our Universe, but its origin is unknown. One of the most attractive possibilities is that DM starts in thermal equilibrium in the early Universe, and its abundance is set once its annihilations become slower than the expansion rate. This framework is insensitive to initial conditions and has the further appeal of tying the DM abundance to its (potentially observable) interactions. 

The most widely considered possibility is that 2-to-2 annihilations to Standard Model (SM) particles set the DM relic density. This is known as the Weakly Interacting Massive Particle (WIMP) paradigm~\cite{Lee:1977ua,Kolb:1990vq,Gondolo:1990dk,Jungman:1995df} and points to DM particles with weak scale masses and cross-sections. This theoretical framework has had considerable impact shaping experimental searches for DM\@.

However it has long been appreciated that simple variations to the cosmology of thermal relics can have dramatic consequences.  In a seminal paper, Ref.~\cite{Griest:1990kh} enumerates three ``exceptions" to thermal relic cosmology: (1) mutual annihilations of multiple species ({\it coannihilations}), (2) annihilations into heaver states ({\it forbidden channels}), and (3) annihilations near a pole in the cross section.  These exceptions lead to phenomenology that can differ significantly from standard WIMPs (see for example Refs.~\cite{ArkaniHamed:2006mb,Tulin:2012uq,Bernal:2015bla,Ibarra:2015nca,Nagata:2015hha,D'Agnolo:2015koa,Baker:2015qna,Delgado:2016umt}), while sharing their appealing theoretical features. 

\begin{figure}[!!!t]
\begin{center}
\includegraphics[width=0.5 \textwidth]{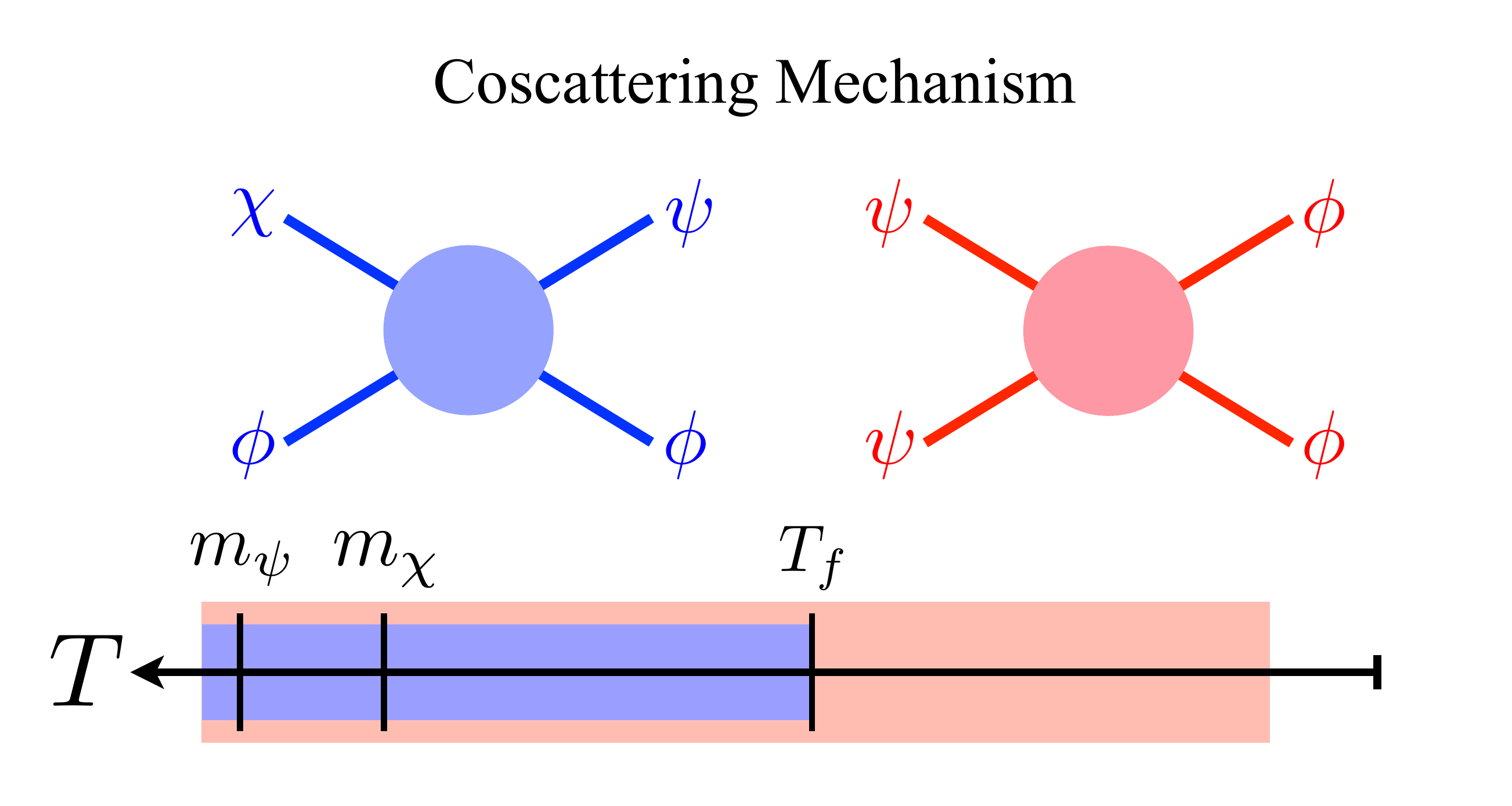}
\end{center}
\vspace{-.3cm}
\caption{\small 
An illustration of the coscattering mechanism for DM freeze-out.  If both diagrams are active, the abundance of DM, $\chi$, decreases through inelastic scattering, $\chi \phi \rightarrow \psi \phi$, followed by annihilations, $\psi \psi \rightarrow \phi \phi$.   Coscattering corresponds to the phase where scattering freezes out before annihilations, setting the DM abundance.
 \label{fig:schema}}
\end{figure}

In this letter, we introduce a fourth exception.  Like Ref.~\cite{Griest:1990kh}, we assume DM begins in thermal equilibrium, has its number diluted through 2-to-2 annihilations, and has a temperature that tracks the photon temperature (for studies that relax at least one of these assumptions see for example Refs.~\cite{Carlson:1992fn,Chung:1998rq,Feng:2003xh, Feng:2008mu,Kaplan:2009ag,Hall:2009bx,D'Eramo:2010ep,Hochberg:2014dra,Kuflik:2015isi,Pappadopulo:2016pkp,Berlin:2016vnh,Farina:2016llk,Dror:2016rxc,Cline:2017tka}).  We consider the presence of two states charged under the symmetry that stabilizes DM: $\chi$ and $\psi$, where $m_\chi < m_\psi$ and $\chi$ is DM\@.  We assume that $\chi$ annihilations are suppressed, and two processes are active:
\begin{enumerate}
\item $\chi/\psi$ interchange: $\chi \phi \leftrightarrow \psi \phi$  (left of Fig.~\ref{fig:schema})
\item $\psi$ annihilations: $\psi \psi \rightarrow \phi \phi$  (right of Fig.~\ref{fig:schema})
\end{enumerate}
where $\phi$ is an unstable state from the thermal bath.  When both processes are in equilibrium, DM number is diluted from  $\chi \rightarrow \psi$ scattering followed by  $\psi \psi$ annihilations. This picture can be generalized to include multiple states $\psi_i, \phi_j$.

In the coannihilation phase, it is assumed that process (2) decouples before process (1), such that the DM abundance is set by the freeze-out of annihilations~\cite{Griest:1990kh}.
We introduce the phase: {\it coscattering}, where process (1) shuts off before process (2), such that the DM abundance is determined by the freeze-out of inelastic scattering.  As we will see, coscattering is generically realized in a large class of models if DM scatters against massive states, $m_\phi \sim m_\chi$.  We note that a similar process was considered within supersymmetry for the special case of an ultralight gluino with a sub-GeV mass, where $\chi$, $\psi$, and $\phi$ were identified with the photino, $R$-hadron, and pion~\cite{Farrar:1995pz,Chung:1997rq}.

\begin{figure}[tb!]
\begin{center}
\includegraphics[width=0.45\textwidth]{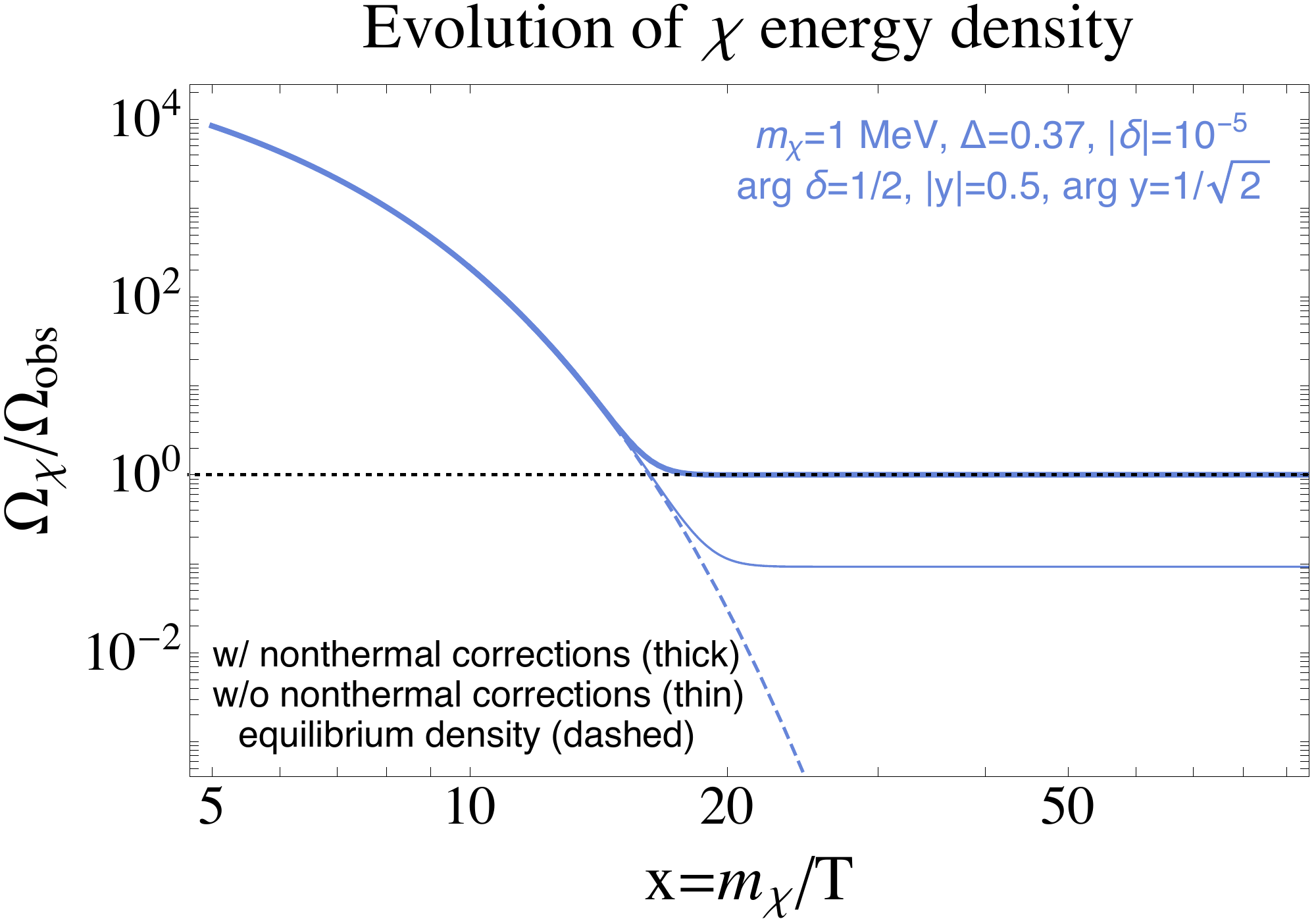}
\end{center}
\vspace{-.3cm}
\caption{Evolution of $\chi$ energy density for coscattering. The thin blue line represent the solution of Eq.~\ref{eq:coscatterBE} where $\chi$ is assumed to be in kinetic equilibrium, while the thick blue line is the solution of the full Boltzmann equation (Eq.~\ref{fullboltzmann}). The dashed blue line represent the equilibrium number density. 
}
\label{fig:Omegavsx}
\end{figure} 

Coscattering leads to unique phenomenology.  As we describe below, the DM abundance has a different parametric form than the WIMP\@.   In order to reproduce the observed abundance, the DM mass is generically much lighter than the weak scale. The DM self-annihilation rate can be arbitrarily small, evading stringent limits from the Cosmic Microwave Background (CMB)~\cite{Ade:2015xua,Slatyer:2015jla,Slatyer:2015kla}.  Although $\chi$ constitutes DM, there is also a relic population of $\psi$ that decay to $\chi$ at late times.  These $\psi$ decays can produce observable distortions to the blackbody spectrum of the CMB. 

The rest of this letter is organized as follows.  We begin by analyzing the relic density of DM produced by coscattering.  We then discuss nontrivial thermal corrections to the abundance, which are further elaborated in the Appendix.  Finally, we 
determine the relic density and experimental constraints in an example model.

\noindent {\bf  Relic Abundance:}\
As above, we consider DM, $\chi$, and a heavier state, $\psi$, that are both charged under the DM stabilizing symmetry.  DM can upscatter into $\psi$ through the coscattering process: $\chi \phi \rightarrow \psi \phi$, where $\phi$ is an unstable state from the thermal bath.  

If $\chi$  and $\psi$ are in kinetic equilibrium (we will relax this assumption in the next section), the evolution of their number densities, $n_{\chi, \psi}$, are determined by the solution to the following system of Boltzmann equations~\cite{Griest:1990kh,Chung:1997rq,Ellis:2015vaa,Nagata:2015hha},
\bea \label{eq:fullBE}
\dot n_i + 3 H n_i &=& - \sum_{j} \Big[n_\phi^{\rm eq} \left< \sigma_{i \rightarrow j} v \right>   \big( n_i - n_i^{\rm eq} \frac{n_j}{n_j^{\rm eq}}  \big) \nonumber \\
&+&    \left< \sigma_{ij} v \right> \left( n_i n_j - n_i^{\rm eq} n_j^{\rm eq} \right)   \Big]
\eea
where $i,j = (\psi, \chi)$, $n_x^{\rm eq}$ denotes the equilibrium Boltzmann distribution, $H$ is the Hubble parameter, and we have assumed that $\phi$ remains in equilibrium.  The first line corresponds to coscattering, $\chi \phi \leftrightarrow \psi \phi$, while the second line corresponds to coannihilations, $\psi \psi, \psi \chi, \chi \chi \rightarrow \phi \phi$.  We have assumed that 2-body decays, $\psi \rightarrow \chi \phi$, are kinematically forbidden: $m_\phi > m_\psi - m_\chi$.  When 2-body decays are active, they typically equilibrate $\psi$ and $\chi$, and then the coscattering diagram does not determine the relic density.  The absence of decays in coscattering is an important difference compared to the light gluino scenario of Refs.~\cite{Farrar:1995pz,Chung:1997rq}, where decays are active.

Coscattering is realized when the following conditions are met: (1)~$\psi \psi \rightarrow \phi \phi$ is in equilibrium, (2)~$\chi \chi, \chi \psi \rightarrow \phi \phi$ can be neglected, and (3) 2-body decays are kinematically forbidden, $m_\phi > m_\psi - m_\chi$.   In this limit, $n_\psi = n_\psi^{\rm eq}$, and the Boltzmann equations simplify,
\be \label{eq:coscatterBE}
\dot n_\chi + 3 H n_\chi = - n_\phi^{\rm eq} \left< \sigma_{\chi \rightarrow \psi} v \right> \left( n_\chi - n_\chi^{\rm eq} \right).
\ee

The solution to Eq.~\ref{eq:coscatterBE} is approximated by taking the DM abundance to be constant after $\chi \leftrightarrow \psi$ decouples, which occurs when
\be \label{eq:SuddenFreeze}
n_\phi^{\rm eq} \left< \sigma_{\chi \rightarrow \psi} v \right> \approx  p \, H,
\ee
where we find that $p \sim 20$ replicates numerical solutions to Eq.~\ref{eq:coscatterBE}.

\begin{figure*}[tb]
\centering
\includegraphics[width=\linewidth]{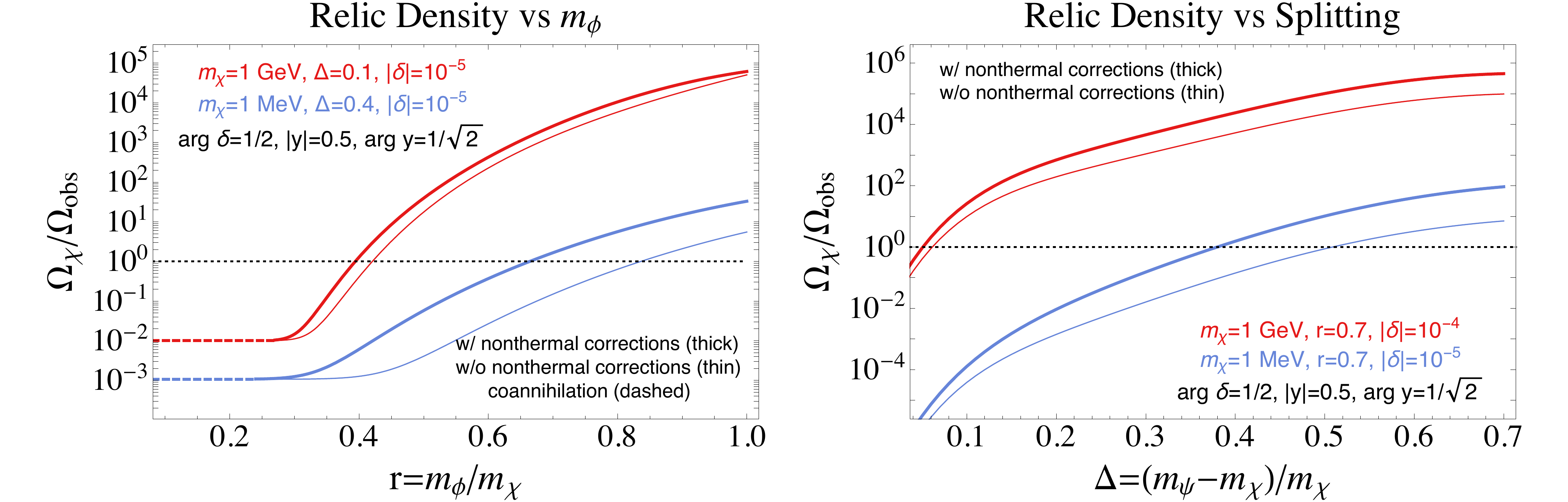}
\caption{ \label{fig:Relic}
The {\it left} side shows the dark matter relic density normalized to its measure value, versus $m_\phi/m_\chi$. 
The plot shows the transition between the coannihilation and coscattering phases, where $\Omega_\chi$ depends exponentially on $m_\phi$. The  {\it right} side shows the relic density normalized to its measure value, versus $\Delta$.
 In both panels the thin solid curves represent the result of the calculation performed assuming kinetic equilibrium for $\chi$, while the thick solid lines are the solution of the full Boltzmann equation (Eq.~\ref{fullboltzmann}).}
\end{figure*}

The $\chi \rightarrow \psi$ scattering is endothermic because $m_\chi < m_\psi$.  The thermally averaged cross section, $\left< \sigma_{\chi \rightarrow \psi} v \right>$, is related to the inverse process, $\psi \rightarrow \chi$, via detailed balance,
\be \label{eq:balance}
 \left< \sigma_{\chi \rightarrow \psi} v \right> =  \frac{n_{\psi}^{\rm eq}}{n_\chi^{\rm eq}}  \left< \sigma_{\psi \rightarrow \chi} v \right> \approx \frac{m_\psi^{3/2}}{m_\chi^{3/2}}  e^{-x \Delta}  \left< \sigma_{\psi \rightarrow \chi} v \right>
\ee
where $x \equiv m_\chi / T$ and $\Delta \equiv (m_\psi - m_\chi)/m_\chi$.  Note that $\left< \sigma_{\chi \rightarrow \psi} v \right>$ is exponentially suppressed in the limit $T \ll m_\psi-m_\chi$.

Using Eq.~\ref{eq:balance} to solve Eq.~\ref{eq:SuddenFreeze}, we find that freeze-out occurs at temperature,
\be \label{eq:freezeTemp}
(r+\Delta) x_f  = 21+\log \left[\frac{(r+r\,\Delta )^{3/2}m_\chi\sigma_{\textrm{inv}}}{p\sqrt{g_*}\, {\textrm{GeV}}\times{\textrm{pb}}}\right]+\log \sqrt x_f
\ee
where $\sigma_{\rm inv} \equiv \left< \sigma_{\psi \rightarrow \chi} v \right>$, $r \equiv m_\phi / m_\chi$, and $g_*$ corresponds to the number of relativistic degrees of freedom at freeze-out. 

Using Eq.~\ref{eq:freezeTemp}, we can estimate the relic density,
\be \label{eq:omega}
\frac{\Omega_\chi}{\Omega_{DM}} \, \approx  \, \frac{0.6\;{\textrm{pb}}}{\sigma_{\rm inv}} \frac{p\,x_f e^{x_f(r+\Delta-1)}}{\sqrt{g_*} r^{3/2} (1+\Delta)^{3/2}}.
\ee
Unlike a WIMP, the abundance (freeze-out temperature) has an exponential (non-logarithmic) sensitivity on the spectrum.

For $r+\Delta > 1$ (i.e. $m_\phi+m_\psi > 2 m_\chi$),  $\sigma_{\rm inv}$ should be exponentially larger than the weak scale in order to reproduce the observed relic density, $\Omega_\chi h^2 \approx 0.12$~\cite{Ade:2015xua}.  This points to DM that is exponentially lighter than the weak scale.  
In the opposite limit, $r+\Delta < 1$, DM cannot be much heavier than the weak scale without violating the requirement that $\psi \psi$ annihilations respect perturbativity and remain in equilibrium until the coscattering process decouples.
It is straightforward to generalize our analysis to multiple states $\psi_i, \phi_j$. 

\noindent {\bf  Departure from kinetic equilibrium:}\
For conventional WIMPs, DM experiences rapid elastic scattering against the thermal bath while annihilations decouple.
Therefore, kinetic decoupling (the departure from Maxwell-Boltzmann phase space distribution) occurs long after chemical decoupling (the freeze-out of number changing interactions); see for example Refs.~\cite{Loeb:2005pm, Bringmann:2006mu, Bringmann:2009vf}.  For coscattering, elastic scattering, $\chi\phi\to\chi\phi$, generically decouples before inelastic scattering, $\chi\phi\to\psi\phi$, because of the small coupling of $\chi$ to the thermal bath. Therefore, $\chi\phi\to\psi\phi$ is responsible for maintaining both chemical and kinetic equilibrium, and its freeze-out brings {\it simultaneous} chemical and kinetic decoupling.  This is an important difference between coscattering and WIMPs and it means that Eq.~\ref{eq:coscatterBE} is not strictly applicable, as it assumes an equilibrium phase space distribution for $\chi$.

In order to correctly treat the departure from kinetic equilibrium, we must solve the full (unintegrated) Boltzmann equation for the time dependence of the momentum space distribution of $\chi$, $f_\chi(p,t)$,
\be\label{fullboltzmann}
\left(\frac{\partial}{\partial t}-H{\bf p}\cdot  \nabla_{\bf p}\right)f_\chi(p,t)=\frac{1}{E}C[f_\chi],
\ee
where $C[f_\chi]$ is the collision operator induced by the coscattering reaction $\chi\phi\to\psi\phi$. $C[f_\chi]$ is a linear function of $f_\chi$.  Therefore, Eq.~\ref{fullboltzmann} is a solvable first-order linear partial differential equation.  We now provide a qualitative sketch of its solution, and we provide more details in the Appendix.  We find that lower momentum modes of $\chi$ decouple earlier than higher momentum modes.  This is because the coscattering process  is endothermic and $\chi$ modes with smaller kinetic energy can only interact with energetic $\phi$ modes from the tail of the Boltzmann distribution with suppressed number density.  Because low momentum modes are more abundant, the final relic abundance of $\chi$ is {\it enhanced} relative to the solution of Eq.~\ref{eq:coscatterBE}.  The size of this thermal correction grows with $\Delta$, which controls the degree of endothermicity of coscattering.  While Eq.~\ref{eq:omega} correctly captures the abundance at the order-of-magnitude level, thermal corrections arising from Eq.~\ref{fullboltzmann} are required for a precise calculation of the abundance, and are included in our numerical results that follow.

\noindent {\bf  An Example Dark Sector:}\
Coscattering is naturally realized within the framework of hidden sector DM~\cite{Goldberg:1986nk,Carlson:1992fn,Strassler:2006im,Finkbeiner:2007kk,Pospelov:2007mp,Feng:2008ya,Feng:2008mu,ArkaniHamed:2008qn,Kaplan:2009ag,Hochberg:2014kqa,Buen-Abad:2015ova}, where $\chi$, $\psi$, and $\phi$ are neutral under the SM gauge group.
We take $\chi, \psi$ to be Majorana fermions, and $\phi$ to be a real scalar, with the following interactions,
\be \label{eq:model}
\mathcal{L} \supset -\frac{m_\chi}{2} \chi^2 - \frac{m_\psi}{2} \psi^2 - \delta m \, \chi \psi - \frac{y}{2} \phi \, \psi^2 + \mathrm{h.c.}~.
\ee
Notice that $\psi$ is active, with Yukawa coupling to $\phi$, while $\chi$ is sterile.  There is a mass mixing, $\delta m$, whose strength is determined by the dimensionless parameter $\delta \equiv \delta m / m_\chi$.  We focus on the small mixing limit, $\delta \ll 1$, where $\psi, \chi$ are approximately mass eigenstates: $n_1 \approx \chi$ and $n_2 \approx \psi$.  Without loss of generality, we take $m_{\chi, \psi}$ to be real and allow generic phases in $y$ and $\delta m$ in order to avoid $p$-wave suppression of the relevant processes. Note that the structure of the interaction of $\chi$ in Eq.~\ref{eq:model} is a natural consequence of a softly broken chiral symmetry.

The annihilation $\psi \psi \rightarrow \phi \phi$ is unsuppressed while $\psi \chi$ and $\chi \chi$ annihilations are suppressed by $\delta^2$ and $\delta^4$, respectively.  The inverse coscattering cross section, $\psi \phi \rightarrow \chi \phi$, which determines the relic density ($\sigma_{\rm inv}$ in Eq.~\ref{eq:omega}) is
$\left< \sigma_{\psi \rightarrow \chi} v \right> \approx f(r) \sqrt{\Delta} \, \frac{y^4 \delta^2}{2 \pi m_\chi^2}$,
where $f(r) \equiv (r^2+r+2)^2/(\sqrt 2(r-2)^2r^{9/2}(r+1)^{7/2})$.  For simplicity, we derive this expression by assuming real $\delta m$ and $y$ and taking the limit $\delta \ll \Delta \ll 1$.

Fig.~\ref{fig:Omegavsx} shows the $\chi$ energy density as a function of $x$. We see that Eq.~\ref{eq:coscatterBE} underestimates the $\chi$ abundance compared to the solution of Eq.~\ref{fullboltzmann}\@. For the parameter choice displayed in Fig.~\ref{fig:Omegavsx}, chemical freeze-out of the coscattering process occurs at $x\approx 20$, while elastic scattering, $\chi\phi\to\chi\phi$, freezes out earlier, $x\approx10$.

\begin{figure}[tb!]
\begin{center}
\includegraphics[width=0.45\textwidth]{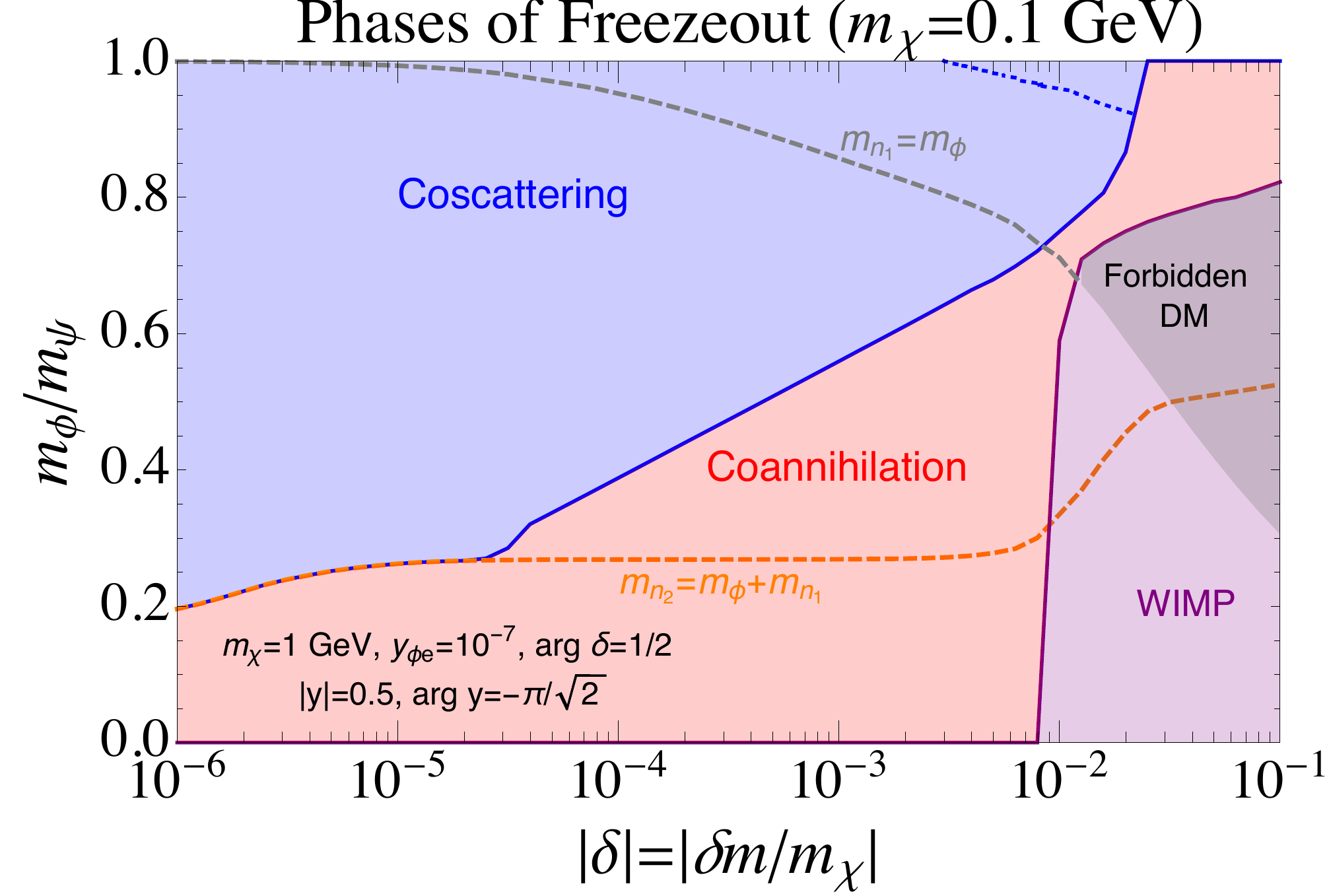}
\end{center}
\vspace{-.3cm}
\caption{Diagram of the different phases of freezeout as a function of ($m_\phi$, $\delta$). Below the dotted blue line, in the coscattering region, elastic scattering, $\chi \phi \rightarrow \chi \phi$, decouples before the coscattering diagram, and thermal effects are important. $\Delta$ is fixed at each point to reproduce the observed relic density.
}
\label{fig:region}
\end{figure} 

Fig.~\ref{fig:Relic} shows how the relic density, $\Omega_{\chi}$, depends on $r \equiv m_\phi/m_\chi$ and $\Delta \equiv (m_\psi - m_\chi)/m_\chi$.  The relic density is exponentially sensitive to these quantities (Eq.~\ref{eq:omega}).
 For the  chosen parameters, the departure from kinetic equilibrium is always relevant. The right of Fig.~\ref{fig:Relic} shows that thermal corrections from Eq.~\ref{fullboltzmann} are enhanced as the splitting $\Delta$ increases.

It is clear from the previous discussion and Fig.~\ref{fig:Relic} that coscattering and coannihilations are closely related~\cite{CoAPrep}.
By varying parameters, any model with coannihilations also realizes coscattering.
Fig.~\ref{fig:region} is the phase diagram, which shows the transition from the coscattering to the coannihilation phase as $\delta$ and $m_\phi$ are varied. Coscattering occurs in the region with small mixing, $\delta \ll 1$, and heavy $\phi$, $m_\phi \sim m_\psi$. This is because the ratio between the coscattering and $\psi\psi\to\phi\phi$ rates scales as
$\sim \delta^2 n_\phi^{\rm eq}/n_\psi^{\rm eq} \sim \delta^2 e^{(m_\psi-m_\phi)/T}$.

For completeness, Fig.~\ref{fig:region} also shows the WIMP phase, where the relic density is set by the freeze-out of $\chi \chi \rightarrow \phi \phi$. It is divided into the conventional case, $m_\chi > m_\phi$, and the forbidden regime~\cite{Griest:1990kh,D'Agnolo:2015koa}, $m_\chi < m_\phi$. 

\begin{figure*}[tb]
\centering
\includegraphics[width=\linewidth]{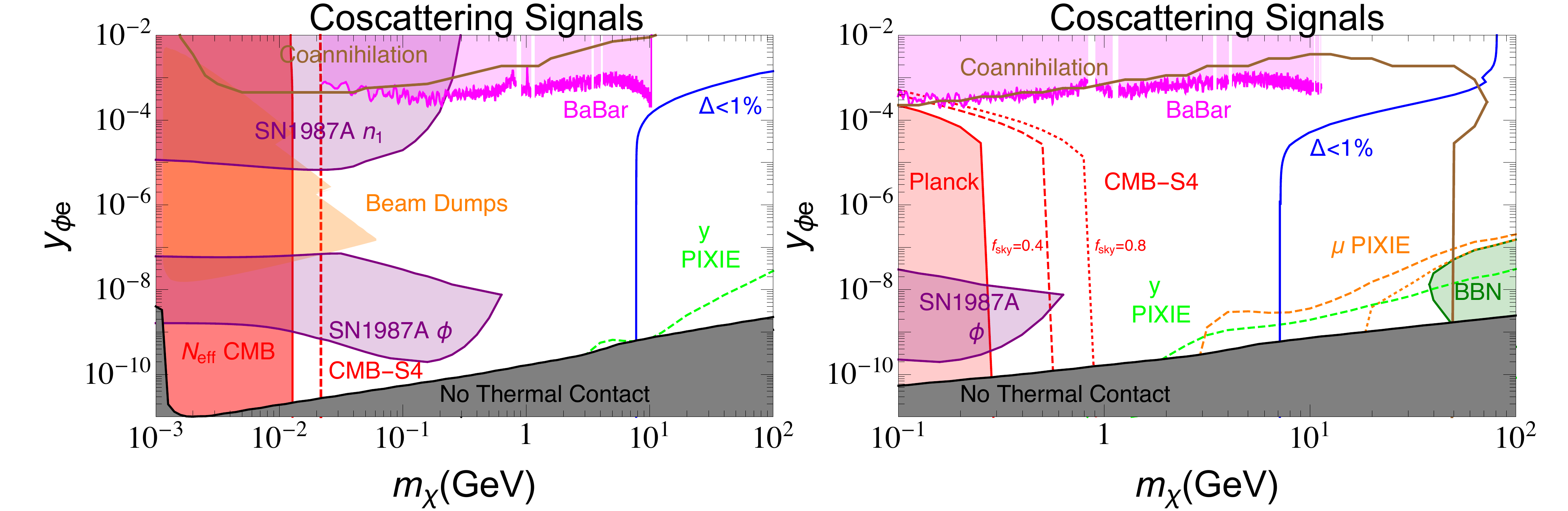}
\caption{ \label{fig:phenopheno}
Phenomenology of the model for a mediator $\phi$ coupling to electrons. Supernova cooling constrains both the direct production of $\phi$ and that of dark matter $n_1$, while we find that $n_2\approx \psi$ is always trapped inside the star. 
The other constraints are described in the main body of the text. The reach of PIXIE corresponds to $\mu < 2.8\times10^{-8}$ and $y<2.4\times10^{-9}$~\cite{Chluba:2013pya, Kogut:2011xw}. The reach including the expected impact of foregrounds, $\mu<9.4\times 10^{-8}$~\cite{Abitbol:2017vwa}, is shown with a dotted line. In the {\it right} panel the reach of CMB Stage-4 measurements on DM annihilations is shown for two different choices of sky coverage 0.4 and 0.8. In the {\it left} panel the remaining model parameters are set to $|y|=1,\,|\delta|=10^{-4},\, \arg y=-i\pi/\sqrt 2,\,\arg\delta=1/2$, and $m_\phi/m_\chi=0.9$. In the {\it right} to $|y|=0.3,\,|\delta|=10^{-3},\, \arg y=-i\pi/\sqrt 2,\,\arg\delta=1/2$, and $m_\phi/m_\chi=0.9$. On both sides, $\Delta$ is fixed at each point to reproduce the observed relic density.
}
\end{figure*}
%

\noindent {\bf  Phenomenology:}\
So far, we have  implicitly assumed that $\phi$ is part of the thermal bath and can decay to other species.  The simplest possibility is that $\phi$ couples to SM particles, leading to experimental signals. In the following, we assume that $\phi$ couples to electrons,
\be
\mathcal{L} \supset - y_{\phi e} \phi \, \bar e e + \mathrm{h.c.} \label{eq:phiele}
\ee
For large enough coupling, $y_{\phi e} \gtrsim 10^{-10}$, the dark sector is in kinetic equilibrium with the SM, implying that the DM temperature tracks the photon temperature. When the coupling becomes too large, $y_{\phi e} \gtrsim 10^{-3}$, dark matter scattering off electrons, $\chi e^{\pm} \to \psi e^{\pm}$, keeps $\chi$ and $\psi$ in equilibrium, bringing the model back into the coannihilation phase.  Coscattering is therefore realized for a wide range of couplings:  $y_{\phi e} \sim 10^{-(3-10)}$.

The various phenomenological constraints are summarized on the right side of Fig.~\ref{fig:phenopheno}. The scalar mediator is constrained by direct production in beam dump experiments~\cite{Bjorken:2009mm, Bjorken:1988as, Riordan:1987aw, Davier:1989wz}, BaBar~\cite{Lees:2014xha}, and supernovae~\cite{Raffelt:1996wa, Burrows:1986me, Burrows:1987zz, Dreiner:2003wh, Dreiner:2013mua, Chang:2016ntp}.  Since $\phi$ couples to electrons but not neutrinos, it modifies their relative temperatures after the weak interactions decouple, changing the effective number of neutrinos, $N_{\rm eff}$~\cite{Boehm:2013jpa}.   We show the current constraints from Planck~\cite{Ade:2015xua} and the projected reach of CMB Stage-4 experiments~\cite{Abazajian:2016yjj}. Planck and CMB Stage-4 measurements are also sensitive to the rate of dark matter annihilations into SM particles~\cite{Ade:2015xua, Slatyer:2015jla, Slatyer:2015kla}, this becomes important at larger values of $\delta$, as shown in the right panel of Fig.~\ref{fig:phenopheno}. 

To conclude this section we discuss a characteristic signal of coscattering. In the coscattering regime, the leading decay of $\psi$ is three-body, $\psi \to \chi e^+ e^-$, and $\psi$ is typically long lived,
\be
\tau_\psi \!\approx 1.2\times10^8 \, \mathrm{s} \left(\frac{10~\mathrm{GeV}}{m_\psi} \right) \left(\frac{10^{-12}}{y_{e \phi} \delta} \right)^2 \left(\frac{0.01}{\Delta} \right)^3 r^4\ .
\ee
These decays can inject energy into CMB photons after the decoupling of double Compton scattering, 
modifying the blackbody spectrum by producing $\mu$ or $y$ distortions~\cite{Hu:1993gc, Chluba:2011hw}.  Current constraints from FIRAS~\cite{Fixsen:1996nj}  do not appear in Fig.~\ref{fig:phenopheno}, but the proposed PIXIE mission~\cite{Kogut:2011xw} has the potential to cover significant new parameter space.  Spectral distortions are typical of coscattering, beyond this particular model realization, because DM upscatters into a heavier state which 
 generically has a trace relic abundance and long lifetime. For the same reasons a fraction of the parameter space is constrained by measurements of light element abundances from BBN~\cite{Poulin:2016anj}.

\noindent {\bf Conclusions:}
In this letter we have introduced the coscattering phase for DM freeze-out.  Coscattering is of broader significance than the example model of Eq.~\ref{eq:model}.  The requirements are (1) mostly sterile DM, $\chi$, with suppressed annihilations; (2) heavier active states, $\psi_i$, with rapid annihilations; and (3) 2-to-2 scatterings against the thermal bath that initially keep DM in equilibrium with the heavier states until these inelastic scatterings decouple and set the DM relic density.  In order to more fully explore the phenomenology of coscattering, it would be interesting to consider more hidden sectors that realize these conditions, and more portals that connect these sectors to the SM. 

\vspace{.3cm}
\begin{acknowledgements}
\noindent {\bf \em Acknowledgements.---}We thank Francesco D'Eramo for collaboration at an earlier stage.  We thank Daniele Alves, Jens Chluba, Rouven Essig, Mariangela Lisanti, Cristina Mondino, Maxim Pospelov, Kris Sigurdson, Tracy Slatyer, and James Wang for helpful discussions.  We thank Bertrand Echenard for providing the numerical limit on a resonance decaying to electrons from Ref.~\cite{Lees:2014xha}.  We thank Lawrence Hall for suggesting the name coscattering. 
RTD is supported by Swiss National Science Foundation contract 200020-169696 and the Sinergia network CRSII2-160814. DP and JTR are supported by NSF CAREER grant PHY-1554858.   RTD and JTR thank the hospitality of the Aspen Center for Physics, which is supported by the NSF grant PHY-1066293.
We thank the Institute for Advanced Study for use of computing facilities.
\end{acknowledgements}
\appendix\label{appendice}
\section{Appendix}

\begin{figure*}[tb!]
\centering
\includegraphics[width=\linewidth]{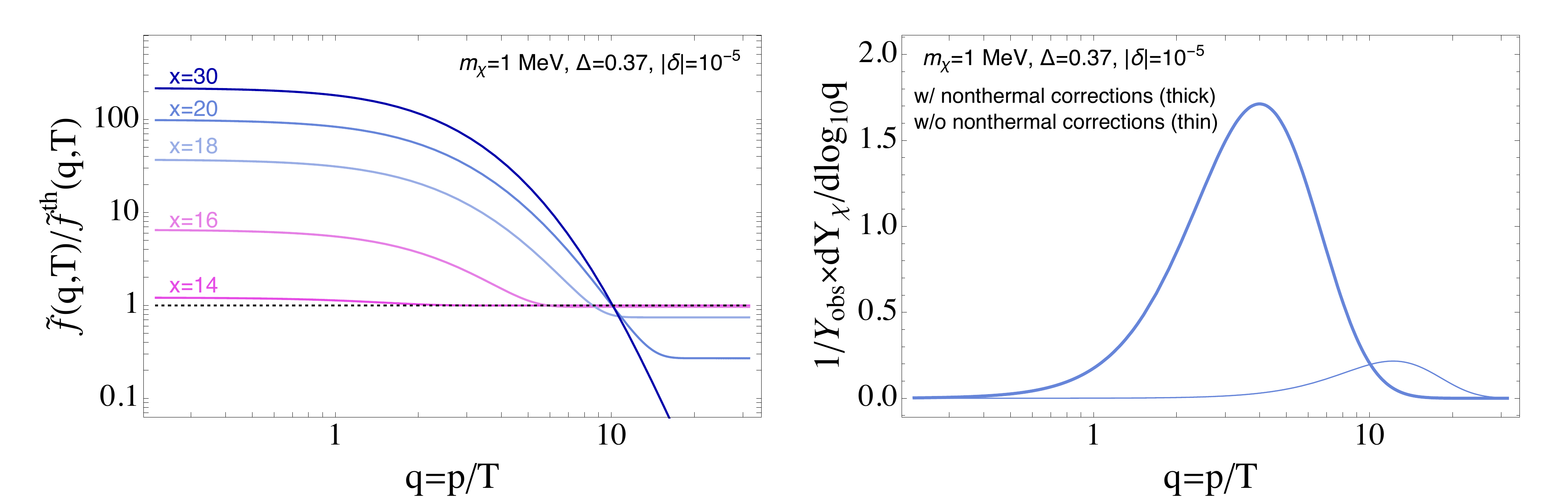}
\caption{ \label{fig:fsol}
The {\it left} plot shows the ratio of the $\chi$ phase space distribution obtained from solving Eq.~\ref{fullboltzmann} to the distribution assuming kinetic equilibrium, see Eq.~\ref{eq:coscatterBE}. The {\it right} plot show a comparison between these two distributions evaluated at late times ($x\sim 30$), when they have become constant.  We plot $1/Y_{\chi\,{\textrm{obs}}} \times d Y_\chi/\log_{10} q$ calculated in terms of $\tilde f$ using Eq.~\ref{densityq}. The final yield is represented by the area below the two curves and matches the observed yield for $m_\chi=1$\,MeV, including thermal corrections. The additional parameters not shown on the plots are fixed to $|y|=0.5$, $\arg y=1/\sqrt2$, and $\arg \delta=1/2$.}
\end{figure*}

In this Appendix, we provide technical details on the solution of the full Boltzmann equation
\be\label{fullboltzmann}
\left(\frac{\partial}{\partial t}-H{\bf p}\cdot  \nabla_{\bf p}\right)f_\chi(p,t)=\frac{1}{E}C[f_\chi],
\ee
for the coscattering process $\chi({\bf p})\phi({\bf k})\to\psi({\bf p}')\phi({\bf k}')$. 

For the above choice of momenta, the collision operator reads
\begin{align}\nn
C[f]=&\frac{1}{2}\int d\Omega_{\bf k} d\Omega_{{\bf k}'} d\Omega_{{\bf p}'} (2\pi)^4\delta^4({\bf p}+{\bf k}-{\bf p}'-{\bf k}')\\
&\overline{|\mathcal M|}^2[f_\phi(k',t)f_\psi(p',t)-f_\phi(k,t)f(p,t)].
\end{align}
The integration measure is the Lorentz invariant phase space 
\be
d\Omega_{\bf p}\equiv \frac{d^3p}{(2\pi)^3 2E(p)}
\ee
and $\overline{|\mathcal M|}^2$ is the matrix element squared averaged over initial quantum numbers and summed over final ones. Assuming that $\phi$ and $\psi$ are in thermal equilibrium we have $f(p,t)=f^{\textrm{eq}}(p,t)=e^{-E(p)/T}$ and we can substitute $f^{\textrm{eq}}_\phi(k',t)f^{\textrm{eq}}_\psi(p',t)=f^{\textrm{eq}}_\phi(k,t)f^{\textrm{eq}}_\chi(p,t)$.
The collision term can thus be rewritten as
\be
\frac{1}{E}C[f]=[f_\chi^{\textrm{eq}}(p,t)-f_\chi(p,t)]\int \frac{d^3 k}{(2\pi)^3}f^{\textrm{eq}}_\phi(k,t)[\sigma v](s)
\ee
where $s=(p_\mu+k_\mu)^2$ and we used the usual definition
\be
[\sigma v](s)=\int d\Omega_{{\bf p}'} d\Omega_{{\bf k}'} (2\pi)^4\delta^4({\bf p}+{\bf k}-{\bf p}'-{\bf k}')\frac{\overline{|\mathcal M|}^2}{4E(p)E(k)}.
\ee
Integrating both sides over the solid angle of ${\bf{p}}$, Eq.~\ref{fullboltzmann} becomes
\be\label{fullboltzmann1}
\left(\frac{\partial}{\partial t}- H p\frac{\partial}{\partial p}\right)f(p,t)=[f_\chi^{\textrm{eq}}(p,t)-f_\chi(p,t)] C(p,t)
\ee
which is thus a linear partial differential equation for $f_\chi$. The function $C$ can be understood as the rate at which different momenta interact with the thermal bath through the coscattering process. 

Changing variables to $q\equiv p\, a$, the comoving momentum, and $a$, the scale factor, the differential operator on the left side of Eq.~\ref{fullboltzmann1} simplifies to:
\be \label{fullboltzmann2}
\left(\frac{\partial}{\partial t}- H p\frac{\partial}{\partial p}\right)f_\chi=H\times a\frac{\partial}{\partial a}\tilde f(q,a),
\ee
where $\tilde f(q,a)\equiv f(q/a,a)$. In these variables, the Boltzmann equation becomes a collection of ordinary linear differential equations, one for each comoving momentum $q$. As a boundary condition for these equations we will use the fact that at an early time, $a_0$, $\chi$ was in kinetic equilibrium with the thermal bath 
\be
\tilde f(q, a_0)=\tilde f^{\textrm{eq}}(q, a_0)=\exp\left(-\sqrt{\frac{q^2}{a_0^2T_0^2}+\frac{m_\chi^2}{T_0^2}}\right)
\ee
where $T_0=T(a_0)$. An explicit solution can thus be written for Eq.~\ref{fullboltzmann2}
\be\label{fullboltzmannsol}
\tilde f_\chi(q,a)=\tilde f^{\textrm{eq}}_\chi(q,a)-\int_{a_0}^a du\frac{d \tilde f^{\textrm{eq}}_\chi(q,u)}{du}e^{-\int_u^a  \frac{\tilde C(q,v)}{v H(v)}dv}.
\ee
The explicit solution immediately implies that $f_\chi\approx f_\chi^{\textrm eq}$ at early times, when $\tilde C/H\gg 1$, as the second term becomes exponentially suppressed. At late times ($T\ll m_\chi$) it is the function $\tilde C/H$ which becomes exponentially suppressed. 

To understand the behavior of the solution we can approximate the exponential by a step function
\be
e^{-\int_u^a  \frac{\tilde C(q,v)}{v H(v)}dv}\approx \Theta(u-a_{\textrm{fo}}(q))
\ee
where $a_{\textrm{fo}}(q)$ is the scale factor at which $\tilde C(q,a)/H(a)$ becomes sufficiently smaller than 1, which can be interpreted as the freeze-out time for the comoving momentum $q$. Plugging this back into Eq.~\ref{fullboltzmannsol} we thus find
\begin{align}\nn
\tilde f_\chi(q,a)&\approx \tilde f^{\textrm{eq}}_\chi(q,a)-\int_{a_0}^a du\frac{d \tilde f^{\textrm{eq}}_\chi(q,u)}{du} \Theta(u-a_{\textrm{fo}}(q))\\
&=\tilde f^{\textrm{eq}}_\chi(q,a_{\textrm{fo}}(q)).
\end{align}
At late times the value of $\tilde f$ for a given comoving momentum is locked to its value at $a_{\textrm{fo}}(q)$.

It is often convenient to use the temperature $T$ as a measure of time, instead of the scale factor $a$. 
Instead of $q=p\times a$, we can take $q=p/T$ as the momentum variable and the solution of Eq.~\ref{fullboltzmann2} becomes the function $\tilde f(q,T)=f(q\, T, T)$. Once  $\tilde f$ is determined, the number density of $\chi$ is calculated by integrating over phase space,
\be\label{densityq}
n_\chi(T)=\int\frac{d^3p}{(2\pi)^3}f(p,T)=T^3\int\frac{d^3q}{(2\pi)^3}\tilde f(q,T).
\ee

The left panel of Fig.~\ref{fig:fsol} shows, for increasing values of $x=m_\chi/T$, the ratio between $\tilde f$ from Eq.~\ref{fullboltzmannsol} and the equilibrium phase space distribution $\tilde f^{{\textrm{th}}}$, including the chemical potential obtained from solving the Boltzmann equation for the number density
\be \label{eq:coscatterBE}
\dot n_\chi + 3 H n_\chi = - n_\phi^{\rm eq} \left< \sigma_{\chi \rightarrow \psi} v \right> \left( n_\chi - n_\chi^{\rm eq} \right).
\ee
As anticipated in the main text, the lower comoving momentum modes decouple earlier from the thermal bath. On the right panel of Fig.~\ref{fig:fsol} we plot $d Y_\chi/d\log_{10} q$ at late times, normalized to the observed yield and calculated using Eq.~\ref{densityq}, with $\tilde f$ either from Eq.~\ref{fullboltzmannsol} or its equilibrium value from Eq.~\ref{eq:coscatterBE}. The plot again shows how the lower momentum modes are enhanced with respect to the thermal case.


\end{document}